\documentclass[prb,twocolumn,superscriptaddress]{revtex4}
\usepackage{amsmath,amssymb}
\usepackage{graphicx}

\begin{document}

\title{Analytic treatment of the precessional (ballistic) contribution
 \\
to the conventional magnetic switching}

\author{Ya.\ B. Bazaliy}
\email{yar@physics.sc.edu}
 \affiliation{Department of Physics and Astronomy, University of
South Carolina, Columbia, SC 29208, USA}
 \affiliation{Institute of Magnetism, National Academy of Science,
Kyiv 03142, Ukraine}

\date{\today}

\begin{abstract}
We consider a switching of the magnetic moment with an easy axis
anisotropy from an ``up'' to a ``down'' direction under the
influence of an external magnetic field. The driving field is
applied parallel to the easy axis and is continuously swept from a
positive to a negative value. In addition, a small constant
perpendicular bias field is present. It is shown that while the
driving field switches the moment in a conventional way, the
perpendicular field creates an admixture of the precessional
(ballistic) switching that speeds up the switching process.
Precessional contribution produces a non-monotonic dependence of the
switching time on the field sweep time with a minimum at a
particular sweep time value. We derive an analytic expressions for
the optimal point, and for the entire dependence of the switching
time on the field sweep time. Our approximation is valid in a wide
parameter range and can be used to engineer and optimize of the
magnetic memory devices.
\end{abstract}

\maketitle

\section{Introduction}
Conventional magnetic switching by an externally applied magnetic
field $\bf H$ is the basis of magnetic recording in hard disk drives
and other devices. The speed at which the moments of the magnetic
bits can be switched between the two easy directions has obvious
implication for the technology performance, setting the limit for
the information writing rate. In general, the magnetic switching
time $\tau_m$ depends on the material parameters, sample size and
shape. Here we are interested in the dependence of $\tau_m$ on the
rate of change of the driving magnetic field. We will call the time
$\tau_h$ required to flip the external magnetic field a ``field
sweep time''. This time is of course finite in any real device used
for magnetic writing. Clearly, very long field sweep times will make
the magnetic switching very slow. One can argue then that, since
$\tau_m$ will decrease with decreasing $\tau_h$, the best case
scenario for switching is the instantaneous flip of the external
field with $\tau_h = 0$. However, it was found numerically
\cite{stankiewicz} that in realistic conditions the function
$\tau_m(\tau_h)$ is not monotonic and has a minimum at a particular
value of the sweep time $\tau_{h}^{*}$. The field sweep time
corresponding to this minimum is optimal and any further decrease of
$\tau_h$ will be counterproductive in terms of the technology
performance. An analytic expression for $\tau_m(\tau_h)$ was
announced in Ref.~\onlinecite{bazaliy:2010}. That paper also
explained the physical reason behind the existence of the switching
time minimum. In the present paper we provide the detailed
derivation of the approximate analytic expressions for the function
$\tau_m(\tau_h)$ and the optimal field sweep time $\tau_{h}^{*}$. We
then discuss the limits of their validity and compare analytic
approximations with the exact numeric results.

Before proceeding to the derivations, we would like to place the
phenomenon under investigation into context. Magnetic switching and
its speed are important technological parameters and were studied by
many authors. We will concentrate on a single-domain magnet
described by a moment $\bf M$ with an easy axis anisotropy energy.
Its switching under a static applied field is described by the
Stoner-Wholfarth astroid \cite{astroid} in the $H$-space
(Fig.~\ref{fig:equilibria}, $\hat z$ is pointing along the easy
axis). If the applied field is changed infinitesimally slowly, and
the thermal fluctuations of the moment can be neglected, magnetic
switching is achieved when the trajectory of the driving field in
the $H$-space crosses the astroid boundary. The switching process
begins at the moment of crossing and requires a finite time that
depends on the crossing point.

If the magnitude and the direction of the applied field are allowed
to change during the magnetic switching process, the number of
switching scenarios becomes infinite. In the most general case one
wants to optimize the dependence ${\bf H}(t)$ so as to minimize the
switching time. Different minimization schemes are discussed
theoretically,\cite{zzsun:2006, sukhov:2009, sukhov:2010} but their
experimental realization is distant as they all require external
fields that change exceedingly rapidly and have carefully controlled
magnitude and direction.

A large body of research is devoted to a restricted case of pulsed
applied fields with fixed direction. The easiest pulse shape to
produce experimentally is a field jump from an initial value ${\bf
H}_i$ to a final value ${\bf H}_f$. The jump can be instantaneous
(step function) or have a certain rise time (smeared step function).
After the jump the field stays at the final value ${\bf H}_f$. The
equilibrium directions of the moment before and after the jump are
described by the Stoner-Wohlfarth picture with ${\bf H} = {\bf H}_i$
and ${\bf H} = {\bf H}_f$ respectively. Before the field jump the
moment resides in the minimum ${\bf M}_i$ of the magnetic energy
corresponding to the initial field. After the field jump the moment
eventually reaches the equilibrium ${\bf M}_f$, corresponding to the
minimum of the energy at the final external field. (We will consider
situations where the energy has only one minimum). The switching
time depends on both the initial and final directions of the moment
or, equivalently, on both ${\bf H}_i$ and ${\bf
H}_f$.\cite{note:adiabatic_switching} This leads to some interesting
properties even for the idealized case of an instantaneous jump. For
example, switching below the Stoner-Wholfarth limit becomes
possible,\cite{he:1994, he:1996, porter:1998, daquino:2005} and is
accompanied by interesting counter-intuitive phenomena. The usual
property of switching time to decrease with with the increasing
magnitude of $H_f$, observed for ${\bf H}_f$ outside of the astroid,
may reverse when ${\bf H}_f$ is inside the astroid.\cite{suess:2002}

If instead of the field jumps one uses pulses with the field
switched from zero to a fixed value for a finite time period, more
options for switching time minimization become available. In the
absence of magnetic anisotropy the fastest 180$^\circ$ flip of a
moment is achieved by applying the field $\bf H$ perpendicular to
${\bf M}_i$. The field is turned on, and the moment starts to
precess around $\bf H$. After a half of the precession period it
reaches the direction ${\bf M}_f = - {\bf M}_i$, at which time the
field has to be switched off. This scheme constitutes the simplest
example of ``precessional'' or ``ballistic'' switching. In the
presence of anisotropy precessional switching becomes more
complicated, but is still possible.\cite{bauer:2000, xiao:2006,
zzsun:2005, wang:2007, horley:2009} The key ingredients of the
precessional switching are extremely short pulses of precise
duration and sufficiently strong field magnitude. Ideally their rise
and fall times should approach zero, but this is very hard to
achieve experimentally as the pulse duration should be of the order
of picoseconds. Precessional switching was observed using the fields
created by pulses of synchrotron radiation \cite{back:1999} or by a
careful pulse shaping using femtosecond lasers,\cite{gerrits:2002}
but is not yet used in applications. The speed of the precessional
switching increases with the field magnitude, but cannot be
increased infinitely due to the breakdown of the ferromagnetic state
of the sample.\cite{tudosa:2004}

Another widely discussed class of switching scenarios is the
application of the high-frequency oscillating field, or an
RF-signal. When the frequency of the RF signal is close to the
eigenfrequency of the magnetic moment, a magnetic resonance occurs
and a continuous precession state is established. The amplitude of
the precession is growing as the strength of the RF signal is
increased. Sufficiently large signal would in principle be able to
increase the amplitude so much as to switch the magnetization from
``up'' to ``down''. Experiments observed an RF-assisted switching
\cite{thirion:2003} happening inside the Stoner-Wohlfarth astroid.
In this case the RF signal helps the external field to drive the
transition. To improve the RF-assisted switching the schemes with
variable (``chirped'') frequency are designed
theoretically.\cite{coffey:1998, garcia:1998, bertotti:2001,
rivkin:2006, zzsun:2006RF, scholz:2008} Their goal is to keep the
frequency equal to the instantaneous, amplitude-dependent resonance
frequency of the magnet. Various instabilities may prevent the
purely RF switching by destroying the coherent single-domain state
of a sample. However, they are supposed to be suppressed in the
magnetic particles with the sizes below $10 \div 20$
nm.\cite{suess:2002,rivkin:2006}

The phenomenon considered in the present paper is different from all
of the above. We consider conventional switching by a field jump,
but concentrate on the switching time $\tau_m$ dependence on the
jump rise time $\tau_h$. The function $\tau_m(\tau_h)$ exhibits an
unexpected minimum which was not discussed in the literature. As it
will be shown in the conclusions, the minimum of $\tau_m(\tau_h)$
results from an admixture of a precessional (ballistic) switching to
the conventional switching, an effect which can be named a
ballistic-assisted switching. Normally, the ballistic contribution
of a small perpendicular field is quenched by the large anisotropy,
but here it is restored by the time-dependence of the switching
field during the rise time of the jump. This type of ballistic
contribution can be observed for infinitesimally small perpendicular
fields, which distinguishes it is from the purely ballistic case
where a finite field comparable to the anisotropy fields is
required. Our analysis also shows that the phenomena of
ballistic-assisted and RF-assisted switching are complimentary. Both
can be treated on equal footing by considering the averaging of the
perturbation field torque during the precession cycle of the moment
in the strong anisotropy field.

\section{Model}
Our treatment is based on the Landau-Lifshitz equation, and does not
take into account thermal fluctuations. We consider a single domain
magnetic bit described by a magnetic moment ${\bf M} = M_0 {\bf n}$,
where ${\bf n}$ is a unit vector. The magnet has an easy anisotropy
axis directed along $z$, and its anisotropy energy is given by $E_a
= -(1/2)K n_z^2$. This anisotropy creates two equilibrium directions
of the moment along $+z$ and $-z$.

The switching field ${\bf H} = H(t) \hat z$ is directed along $z$ as
well. It favors the $+z$ direction for $H > 0$ and $-z$ for $H < 0$.
For large enough magnitudes, $|H| > K/M_0$, there is no equilibrium
in the direction opposite to the field.

When the external field is pointing along the easy axis direction,
magnetic switching relies on the fluctuations near the equilibrium
position. Without them a moment pointing exactly along $+z$ will not
be switched by any amount of negative applied field. Instead it will
remain at the point of unstable equilibrium, until an initial
fluctuation occurs and then grows with time. The switching time in
this case strongly depends on the fluctuation magnitude
$\delta\theta$, being infinite for $\delta\theta =
0$.\cite{kikuchi:1956, uesaka:2002} In order to model the required
initial fluctuation we apply a small bias field ${\bf H}_{\perp} =
H_{\perp} \hat x$ perpendicular to the easy axis. This field is set
to be constant in time. It creates a controlled deviation of the
initial magnetization from the field direction and makes the problem
well defined.

The switching dynamics is described by the Landau-Lifshitz-Gilbert
(LLG) equation.
$$
\dot{\bf M} = -\gamma \left[\frac{\partial E}{\partial {\bf M}}
\times {\bf M} \right] + \frac{\alpha}{M_0} [{\bf M} \times \dot{\bf
M}] \ ,
$$
where $\gamma$ is the gyromagnetic ratio, $E = E_a - ({\bf H} + {\bf
H}_{\perp})\cdot {\bf M}$ is the magnetic energy, and $\alpha$ is
the Gilbert damping constant. In terms of the unit vector $\bf n$ it
reads
\begin{equation}
\dot{\bf n} = -\left[\frac{\partial \varepsilon}{\partial {\bf n}}
\times {\bf n} \right] + \alpha [{\bf n} \times \dot{\bf n}]
\end{equation}
with $\varepsilon = \gamma E / M_0$. Using the spherical angles
($\theta, \phi$), defined so that ${\bf n} = \{n_x, n_y, n_z \} = \{
\sin\theta \cos\phi, \sin\theta\sin\phi, \cos\theta \}$, one gets a
system of equations
\begin{eqnarray}
 \label{LLGtheta}
(1+\alpha^2) \dot{\theta} &=&
  - \frac{1}{\sin\theta}\frac{\partial \varepsilon}{\partial\phi} - \alpha \frac{\partial
  \varepsilon}{\partial\theta} \ ,
 \\
 \label{LLGphi}
(1+\alpha^2) \dot{\phi} &=&
 \frac{1}{\sin\theta}\frac{\partial \varepsilon}{\partial\theta} -
 \frac{\alpha}{\sin^2\theta} \frac{\partial \varepsilon}{\partial\phi} \ .
\end{eqnarray}
Actual materials are characterized by $\alpha \ll 1$ and we will
always calculate up to the linear terms in $\alpha$, e.g., $1 +
\alpha^2 \approx 1$. In our case the energy has the form
\begin{eqnarray}\label{eq:uniaxial_total_energy_1}
\varepsilon(\theta,\phi) &=& -\frac{\omega_0}{2} n_z^2
 - h(t) n_z - h_{\perp} n_x =
 \\
 \nonumber
 &=& -\frac{\omega_0}{2}\cos^2\theta - h(t) \cos\theta - h_{\perp}
\sin\theta\cos\phi \ ,
\end{eqnarray}
where $\omega_0 = \gamma K/M_0$, $h = \gamma H$, $h_{\perp} = \gamma
H_{\perp}$. The smallness of the bias field is ensured by the
condition $h_{\perp} \ll \omega_0$.

\begin{figure}[t]
\center
\includegraphics[width = 0.4\textwidth]{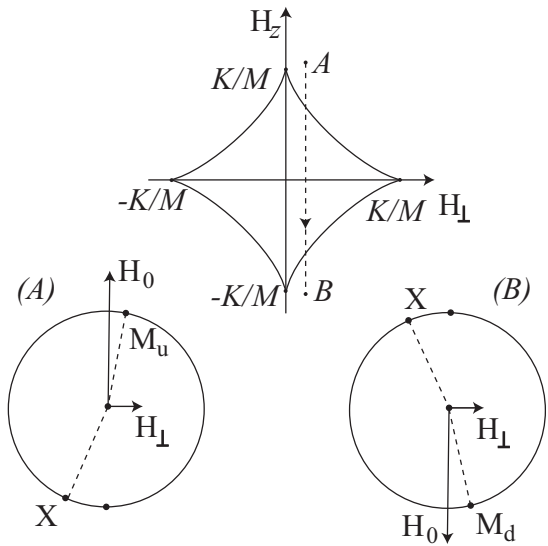}
\caption{Top: Stoner-Wohlfarth astroid in the $H$-space and the
trajectory of the magnetic field change (dashed line) during the
rise time of the pulse with axial field sweep from $+H_0$ to $-H_0$
at a constant perpendicular bias $H_{\perp}$. The lower panels (A)
and (B) correspond to the initial and final directions of the field.
They show the ``up'' and ``down'' minimum points $M_u$, $M_d$, and
maximum points $X$ of the magnetic energy. }
 \label{fig:equilibria}
\end{figure}

The field sweep is assumed to be linear in time and given by the
expressions
\begin{eqnarray}
 \nonumber
h(t) &=& + h_0 \ , \quad (t < 0)
 \\
 \label{eq:field_sweep_form}
h(t) &=& h_0 \left( 1 - \frac{2 t}{\tau_h} \right) \ , \quad (0 < t
< \tau_h)
 \\
 \nonumber
h(t) &=& - h_0 \ , \quad (t > \tau_h) \ .
\end{eqnarray}
The trajectory of the magnetic field change and the positions of the
minima and maxima of the magnetic energy for the initial and final
field orientations are shown in Fig.~\ref{fig:equilibria}. We start
with the positive field $h = +h_0 > \omega_0$ which guarantees that
the magnet is initially pointing close to the $+z$ direction. This
state will be called an ``up-equilibrium''. At the end of the
reversal the moment reaches the ``down-equilibrium", corresponding
to $h = -h_0$. The initial and final directions of the moment are
determined from the conditions $\partial\varepsilon/\partial\phi =
0$, $\partial\varepsilon/\partial\theta = 0$ with $h = \pm h_0$.
This gives $\phi = 0$ and an equation for $\theta$ reading
$$
\omega_0 \sin\theta\cos\theta \pm h_0 \sin\theta -
h_{\perp}\cos\theta = 0 \ .
$$
In the limit $h_{\perp} \ll \omega_0$ one finds the following
approximations for the values of the polar angles in the up- and
down-equilibria
\begin{equation}
\theta_{u} \approx  \frac{h_{\perp}}{\omega_0 + h_0} \ ,
 \quad
\theta_{d} \approx \pi - \frac{h_{\perp}}{\omega_0 + h_0} \ .
\end{equation}

As the field is swept from positive to negative values, the
up-equilibrium disappears and the magnetic moment starts to move in
a spiral fashion towards the down-equilibrium, approaching it
exponentially. To define a finite switching time we have to
introduce a provisional cut-off angle $\theta_{sw}$ and calculate
the time it takes to reach $\theta_{sw}$ during the switching
process. The remaining distance from $\theta_{sw}$ to $\theta_d$
takes extra time, but this time interval does not depend on the
field sweep time since in the regimes studied in our paper $\tau_h <
\tau_m$, and the remaining motion happens at a constant external
field. We use the commonly adopted \cite{stankiewicz, suess:2002,
he:1994, porter:1998} value of $\theta_{sw} = \pi/2$.

\section{Numeric results: non-monotonic dependence of the switching time}
\begin{figure}[t]
\center
\includegraphics[width=0.45\textwidth]{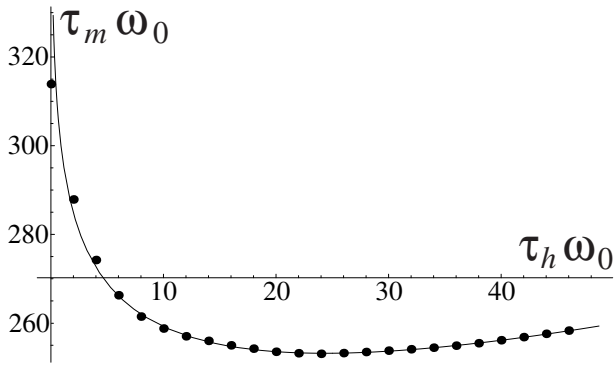}
\caption{Switching time as a function of field sweep time (time is
measured int he units of $\omega_0^{-1}$). Parameter values are
$\alpha = 0.01$, $h_0 = 3.5 \ \omega_0$, $h_{\perp} = 0.001 \
\omega_0$. The thin solid line is the approximate expression derived
in this paper.}
 \label{fig:tsw_numeric}
\end{figure}

The LLG equation can be easily solved numerically and the switching
time dependence $\tau_m(\tau_{h})$ can be obtained.
Fig.~\ref{fig:tsw_numeric} shows the results of such modeling for a
particular parameter set (see figure caption). The minimum of
$\tau_m$ is clearly observed.

We will show in the subsequent sections that the switching time can
be approximated by an expression
\begin{eqnarray*}\label{eq:tau_m_preview}
\tau_m &=& \frac{3 h_0 + \omega_0}{4 h_0} \tau_h + \frac{\ln[h_0/\pi
h_{\perp}^2 \tau_h]}{2 \alpha(h_0 -\omega_0)} + \tau_R
(\alpha,h_0,\omega_0)
\end{eqnarray*}
where $\tau_R$ is independent of $\tau_h$ and $h_{\perp}$. As one
can see in Fig.~\ref{fig:tsw_numeric}  the correspondence between
the actual (numeric) and approximate curves is quite good and
reproduces the minimum of $\tau_m$.

A note on the numeric calculation is due here. It is more convenient
to follow the time dependence of the total energy $\varepsilon(t)$,
than that of $\theta(t)$. The reason for that is as follows. In the
regime considered here the switching time and field sweep time
satisfy $\tau_m > \tau_h$ and the switching threshold $\theta =
\pi/2$ is reached when the external field is already
time-independent, $h = -h_0$. According to the LLG equation, at
constant external field the time derivative of energy is strictly
negative
$$
\dot\varepsilon(t) = - \alpha \dot{\bf n}^2 < 0 \ ,
$$
and $\varepsilon(t)$ is a strictly decreasing function of time. This
property greatly simplifies the solution of the equation
$\varepsilon(\tau_m) = \varepsilon_{sw}$, where $\varepsilon_{sw}$
is the new cut-off, introduced instead of $\theta_{sw}$. Note that
in contrast to $\varepsilon(t)$ the time dependence of $\theta(t)$
is non-monotonic. One can easily understand that by recalling that
in the $\alpha \ll 1$ limit the spiral motion of the moment
approximately follows the equipotential lines. Due to the presence
of the bias field $h_{\perp}$ the latter differ from the $\theta =
{\rm const}$ lines, leading to the oscillations of $\theta(t)$ in
time. In other words, the fact that $\varepsilon(\theta,\phi)$
depends on both spherical angles makes the cut-offs $\theta_{sw}$
and $\varepsilon_{sw}$ not completely equivalent. Nevertheless, they
serve the same purpose with the latter being a more convenient
choice. We adopt the cut-off value $\varepsilon_{sw} =
\varepsilon(\pi/2, \pi/2) = 0$, closest to the original
definition.\cite{stankiewicz}

\section{Analytic approximation for the switching time}
The switching process consists of two stages. The first stage is the
field sweep time interval, $0 < t < \tau_h$. The second stage is the
motion in the constant field for the time interval $\tau_h < t <
\tau_m$. Below we use two different approximations to find the
magnetic dynamics in each stage.

\subsection{First stage}
The idea for the first stage approximation is to assume that the
deviation of $\bf n$ from $+z$ is small. The rational for that is
provided by the following argument. The initial position of the
moment is given by $\theta_u \approx h_{\perp}/(\omega_0 + h_0) \ll
1$, i.e., is very close to $+z$. We then assume that proximity to
the $+z$ holds throughout the first stage if the sweep time is not
too long. The precise condition imposed by this assumption on
$\tau_h$ is not clear at this point but will be obtained after we do
the calculations.

According to the above, we linearize the LLG equation near the $+z$
point. In the linearized equation the unknowns are the two
projections $(n_x,n_y)$ of the unit vector. Both are small for ${\bf
n}$ close to $+z$. One gets a linear system
\begin{eqnarray*}
\dot n_x &=& - \alpha (\omega_0+h) n_x - (\omega_0+h) n_y  + \alpha
h_{\perp}
 \\
\dot n_y &=& = (\omega_0+h) n_x - \alpha (\omega_0+h) n_y- h_{\perp}
\ .
\end{eqnarray*}
Introducing a notation $\omega(t) = \omega_0 + h(t)$ we rewrite it
as
\begin{equation}
\left( \begin{array}{c}
 \dot n_x \\ \dot n_y \end{array} \right) =
 \left| \begin{array}{cc}
 -\alpha\omega(t) & -\omega(t)
 \\
 \omega(t) & -\alpha\omega(t)
 \end{array}\right|
 \left( \begin{array}{c}
 n_x \\ n_y \end{array} \right) +
 \left( \begin{array}{c}
 \alpha h_{\perp} \\ -h_{\perp} \end{array} \right)
\end{equation}
The matrix on the right hand side can be diagonalized by changing
variables to $\xi = n_x + i n_y$, $\eta = n_x - i n_y$. Using them
we get two decoupled equations
\begin{eqnarray*}
\dot\xi &=& (i-\alpha)\big[\omega(t) \xi - h_{\perp} \big] \ ,
 \\
\dot\eta &=& -(i+\alpha)\big[\omega(t) \eta - h_{\perp} \big] \ ,
\end{eqnarray*}
which turn out to be complex conjugates of each other. Consequently,
we can solve either one of them. Denoting $\mu = i - \alpha$, we
search for the solution of the first equation in the form
$$
\xi(t) = A(t) e^{\mu \int^t_0 \omega(s) ds} \ .
$$
For future notation we define a phase function $\varphi(t) =
\int^t_0 \omega(s) ds$. The solution is found to be
$$
\xi(t) = \xi(0) e^{\mu \varphi(t)}
  - \mu h_{\perp} e^{\mu \varphi(t)} \int^t_0  e^{-\mu \varphi(u)}du
  \ .
$$
Going back to $(n_x,n_y)$ we obtain after the necessary algebraic
transformations a solution
\begin{eqnarray}
 \nonumber
n_x &=& e^{-\alpha\varphi} \left\{
 n_{x0}\cos\varphi - n_{y0}\sin\varphi - \right.
 \\
 \nonumber
 &-& \left. h_{\perp}
   \left[
   (S-\alpha C)]\cos\varphi - (C + \alpha S)]\sin\varphi
   \right] \right\} \ ,
 \\
 \label{eq:nxny_exact}
n_y &=& e^{-\alpha\varphi} \left\{
 n_{x0}\sin\varphi + n_{y0}\cos\varphi - \right.
 \\
 \nonumber
 &-& \left. h_{\perp}
   \left[
   (S-\alpha C)]\sin\varphi + (C + \alpha S)]\cos\varphi
   \right] \right\} \ ,
\end{eqnarray}
where we have defined
\begin{eqnarray}
 \label{eq:defSandC}
S(t) &=& \int_0^t e^{\alpha\varphi(s)} \sin\phi(s) ds \ ,
 \\
 \nonumber
C(t) &=& \int_0^t e^{\alpha\varphi(s)} \cos\phi(s) ds \ .
\end{eqnarray}
In our case the initial conditions are given by
\begin{equation}\label{eq:initial_conditions}
n_{x0} = \frac{h_{\perp}}{\omega_0 + h}, \quad n_{y0} = 0 \ ,
\end{equation}
thus both projections $n_x$ and $n_y$ are proportional to
$h_{\perp}$ and we should be able to satisfy the assumption of small
deviation from the origin for sufficiently small bias field. Below
we will calculate how small should $h_{\perp}$ be to ensure small
deviations in Stage I.

For the linear field sweep (\ref{eq:field_sweep_form}) the phase
$\varphi(t)$ is a quadratic function
$$
\varphi = \int_0^t \left(
  \omega_0 + h_0\left(1-\frac{2 s}{\tau_h}\right)
  \right) ds =
  (\omega_0 + h_0) t - h_0 \frac{t^2}{\tau_h} \ .
$$
In this case the integrals (\ref{eq:defSandC}) can be found exactly
and expressed through the error function of complex argument
(Appendix~\ref{appendixI}).

Here we will consider a useful approximation valid in a large region
of parameters. The phase $\varphi(t)$ has one maximum on the
interval $[0,\tau_h]$ at the point $t_m = \tau_h(\omega_0 +
h_0)/2h_0$ (recall that $\omega_0 < h_0$). The presence of a maximum
means that the integrals for $C(\tau_h)$ and $S(\tau_h)$ can be
approximated by a steepest descent (stationary phase) method in the
case of a large change of $\varphi$ on the integration interval
$[0,\tau_h]$. This certainly requires the inequality $\omega_0
\tau_h \gg 1$ to hold but, as it turns out below, sometimes an even
stronger condition is needed. The steepest descent calculation is
performed in Appendix~\ref{appendixI} and gives an approximation
\begin{eqnarray}
 \nonumber
&& S = e^{\alpha\phi_m} \sqrt{\frac{\pi\tau_h}{h_0} } \left[
 \sin\left(\phi_m - \frac{\pi}{4}\right)
 + \frac{\alpha}{2} \cos\left(\phi_m - \frac{\pi}{4}\right)
 \right] \ ,
 \\ \nonumber
&& C = e^{\alpha\phi_m} \sqrt{\frac{\pi\tau_h}{h_0} } \left[
 \cos\left(\phi_m - \frac{\pi}{4}\right)
 - \frac{\alpha}{2} \sin\left(\phi_m - \frac{\pi}{4}\right)
 \right] \ ,
 \\ \label{eq:stationary_phase_SandC}
&& \varphi_m  \equiv \varphi(t_m) = \frac{(\omega_0 + h_0)^2}{4 h_0}
\tau_h \ .
\end{eqnarray}
Substituting this into (\ref{eq:nxny_exact}), using $\varphi(\tau_h)
= \omega_0\tau_h$ and performing the calculations we find
\begin{eqnarray}
 \nonumber
n_x(\tau_h) &=& e^{-\alpha\omega_0\tau_h}
 \frac{h_{\perp}}{\omega_0 + h_0} \cos\omega_0 t -
 \\
 \nonumber
& - & e^{\alpha \Delta\varphi} h_{\perp}
 \sqrt{\frac{\pi \tau_h}{h_0}}
 (\sin \overline{\Delta\varphi}
 - \frac{\alpha}{2}\cos\overline{\Delta\varphi}) \ ,
 \\ \label{eq:nxny_at_tauh}
n_y(\tau_h) &=& e^{-\alpha\omega_0\tau_h}
 \frac{h_{\perp}}{\omega_0 + h_0} \sin\omega_0 t -
 \\
 \nonumber
&& - e^{\alpha \Delta\varphi} h_{\perp}
 \sqrt{\frac{\pi \tau_h}{h_0}}
 (\cos\overline{\Delta\varphi} + \frac{\alpha}{2}
 \sin\overline{\Delta\varphi}) \ ,
\end{eqnarray}
where we have defined
\begin{eqnarray}
 \label{eq:Delta_phi}
\Delta\varphi &=& \varphi(t_m) - \varphi(\tau_h) =
 \frac{(\omega_0 - h_0)^2}{4 h_0} \tau_h \ ,
 \\ \nonumber
 \overline{\Delta\varphi} &=& \Delta\varphi - \frac{\pi}{4} \ .
\end{eqnarray}
We observe  that in both formulae the first term on the right hand
side is initially small and further decreases as a function of
$\tau_h$, while the second term increases with $\tau_h$.
Therefore the condition for small deviations can be formulated as
the smallness of the second term
\begin{equation*}
e^{\alpha \Delta\varphi} h_{\perp}
 \sqrt{\frac{\pi \tau_h}{h_0}} \ll 1 \ .
\end{equation*}
Explicitly separating the product $\omega_0 \tau_h$, we can write
the condition on the bias field
\begin{equation}\label{eq:stageI_validity_condition}
\frac{h_{\perp}}{\omega_0} \ll
 \sqrt{\frac{h_0}{\pi\omega_0}} \sqrt{\frac{1}{\omega_0 \tau_h}}
 \exp\left[-
 \alpha \frac{(\omega_0 - h_0)^2}{4 h_0 \omega_0}
(\omega_0\tau_h) \right] \ ,
\end{equation}
which will guarantee the validity of the small deviations
assumption. Additional inequalities
(\ref{eq:appendix_steepest_descent_conditions}) enabling the
approximation (\ref{eq:stationary_phase_SandC}) are listed in
Appendix~\ref{appendixI} and have to be satisfied as well. We will
return to their discussion in Sec.~\ref{sec:validity}.

\subsection{Second Stage}
During the Stage II the external magnetic field is constant, ${\bf
H} = - H_0 \hat z + H_{\perp} \hat x$. The action of the bias field
$H_{\perp}$ can be viewed as a perturbation of the axially symmetric
problem with $H_{\perp} = 0$ and $H = -H_0$. For the unperturbed
problem the switching time is a known \cite{kikuchi:1956,
uesaka:2002} as a function of $\theta_{in}$, the angular deviation
of the magnetization from the easy axis at the beginning of
Stage~II.

To find the perturbation corrections to the axially symmetric
problem we employ the method of deriving an approximate differential
equation for the total energy $\varepsilon$ in the limit of small
Gilbert damping constant (see, e.g., Ref.~\onlinecite{bonin:2007}).
In the $\alpha \to 0$ limit the motion of the moment ${\bf M}(t)$
can be viewed as a fast precession along the $\varepsilon = $ const
lines and a slow motion from one equipotential orbit to the next one
nearby. Up to the linear terms in $\alpha$ the change of energy upon
one precession cycle around an orbit is given by
\begin{equation}\label{eq:dissipation_integral}
\Delta \varepsilon =
 -\alpha
  \oint_{\Gamma} \left| \frac{\partial \varepsilon}{\partial {\bf n}} \right| dn
  = - \alpha f_1(\varepsilon) \ ,
\end{equation}
and the period of this cycle is
\begin{equation}\label{eq:period_integral}
T  = \oint_{\Gamma} \frac{dn}{\left|\partial \varepsilon/ \partial
{\bf n}
 \right|} = f_2(\varepsilon) \ ,
\end{equation}
where the integrals are taken along the constant energy orbit
$\Gamma(\varepsilon)$ on the unit sphere. In this approximation the
differential equation for $\varepsilon(t)$ reads \cite{bonin:2007}
\begin{equation}\label{eq:genericEvst}
\frac{d\varepsilon}{dt} = \frac{\Delta \varepsilon}{T} = -
\alpha\frac{f_1(\varepsilon)}{f_2(\varepsilon)} = -\alpha
\psi(\varepsilon) \ .
\end{equation}
Its solution is given by
$$
t = -\frac{1}{\alpha} \int_{\varepsilon_1}^{\varepsilon_2}
\frac{d\varepsilon}{\psi(\varepsilon)} \ ,
$$
and determines the time required to move from the orbit with energy
$\varepsilon = \varepsilon_1$ to the orbit with $\varepsilon =
\varepsilon_2$.

\begin{figure}[t]
\center
\includegraphics[width=0.3\textwidth]{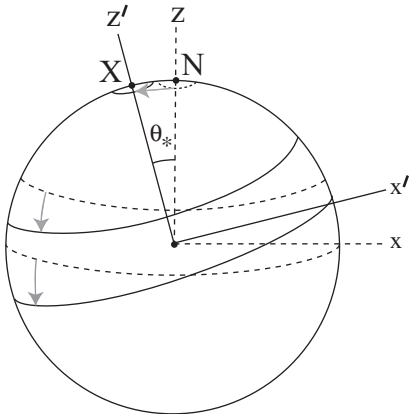}
\caption{Rotated reference frame and the definition of $\theta_*$.
Unperturbed equipotential orbits are shown in dashed lines and the
actual orbits given by the solid lines. Grey arrows show the
transformation of the representative orbits as the bias field
$h_{\perp}$ is turned on. Point $X$ is the energy maximum position
in the presence of bias field.}
 \label{fig:rotated_frame}
\end{figure}

To find the integrals (\ref{eq:dissipation_integral}) and
(\ref{eq:period_integral}), we have to find the orbits
$\Gamma(\varepsilon)$. When $h_{\perp}$ is small, one can expect
that the equipotential orbits will be close to those in the
unperturbed case with $h_{\perp} = 0$. The latter are the circles of
constant polar angle. Indeed, in the absence of the bias field
$\varepsilon = \varepsilon_0(\theta)$ and for any given energy the
polar angle is given by an inverse function $\theta =
\theta_0(\varepsilon)$. This statement is true most of the time,
however important exceptions exist. As shown in
Fig.~\ref{fig:rotated_frame}, the orbits are indeed relatively close
near the equator. But near the North pole $N$ of the sphere the
orbits of the perturbed energy are small circles around the maximum
point $X$, while the original orbits are small circles around $N$.
They are not relatively close in the sense that the perturbation is
larger than the orbit size.

One can remedy the situation by switching to a coordinate system
with the $z'$ axis going through the maximum point $X$. The new
coordinate system $(x',y',z')$ is rotated with respect to the
original $(x,y,z)$ system by an angle $\theta_*$ around the fixed
$y' = y$ axis (see Fig.~\ref{fig:rotated_frame}). The spherical
angles of the new system will be denoted by $\theta'$ and $\phi'$.
The advantage of the rotated system comes form the fact that the
perturbed orbits are close to $\theta' = {\rm const}$ circles
everywhere, except in the vicinity of $\theta' = \pi$. For the
region of interest $0 < \theta' \leq \pi/2$ it is possible to define
the shape of the orbit $\Gamma(\varepsilon)$ as a perturbation
$\theta'(\phi',\varepsilon) = \theta_0(\varepsilon) + h_{\perp}
\delta\theta(\phi',\varepsilon)$ around the circles of constant
$\theta'$.

To prove the geometrically intuitive statement of the preceding
paragraph one has to invert the equation $\varepsilon(\theta',\phi')
= \varepsilon$. The form of energy function in the new spherical
angles if calculated in the
Appendix~\ref{appendix_energy_rotated_coordinates}. Up to the first
order in $h_{\perp}$ we find
\begin{equation}\label{eq:energy_prime_approx}
\varepsilon =  \varepsilon_0(\theta')
 + \beta \varepsilon_1(\theta',\phi')
 + \ldots
\end{equation}
where the small parameter is
\begin{equation}\label{eq:defbeta}
\beta = \frac{h_{\perp}}{h_0 - \omega_0}
\end{equation}
and
\begin{eqnarray}
 \nonumber
\varepsilon_0(\theta') &=& -\frac{\omega_0}{2} \cos^2\theta'
 + h_0 \cos\theta'
 \\ \label{eq:varepsilon1}
\varepsilon_1(\theta') &=& \omega_0 (1 -
\cos\theta')\sin\theta'\cos\phi'
\end{eqnarray}
As expected, the zeroth order term is given by the unperturbed
energy as a function of the new polar angle $\theta'$. The same
would have happened in the original coordinates, but there is an
important difference between the new and original coordinates, which
manifests itself in the behavior of $\varepsilon_1$ at small values
of $\theta'$. Note that the second term in
(\ref{eq:energy_prime_approx}) grows with $\theta'$ slower than the
first one: $\varepsilon_1 \sim {\theta'}^3$, while $\varepsilon_0 -
\varepsilon_0(0) \sim {\theta'}^2$. As a result, the inequality
$\beta\varepsilon_1 \ll \varepsilon_0 - \varepsilon_0(0)$ holds
uniformly in $\theta'$. We will see in a moment that it is precisely
this uniformity that is important. In the original coordinates as
$\theta \to 0$ at an arbitrarily small but fixed $\beta$ the second
term exceeds the first one, and the uniformity is violated.

The orbit equation $\theta' = \theta'(\phi',\varepsilon)$ is found
form the constant energy condition
$$
\varepsilon_0(\theta') + \beta (1 - \cos\theta')\sin\theta'\cos\phi'
= \varepsilon \ ,
$$
which has to be solved for $\theta'$. The solution is searched in
the form of a power series in $\beta$
$$
\theta' = \theta_0(\varepsilon) + \beta \theta_1(\phi',\varepsilon)
+ \ldots
$$
where $\theta_0(\varepsilon)$ is the inverse function of
$\varepsilon_0(\theta)$, as was already discussed above. Up to the
first order in $\beta$ we find
\begin{equation}\label{eq:orbit_shape}
 \theta' = \theta_0(\varepsilon) - \beta
\frac{\varepsilon_1(\theta_0(\varepsilon),\phi')}
{(d\varepsilon_0/d\theta)|_{\theta = \theta_0(\varepsilon)}} +
\ldots
\end{equation}
In the ``dangerous'' limit of small $\theta_0(\varepsilon)$ near the
North pole the second term in (\ref{eq:orbit_shape}) is proportional
to ${{\theta_0}^2(\varepsilon)}$ thus being a small correction. Due
to $\beta \ll 1$ it remains a small correction for the energy values
up to near the equator $\varepsilon \approx 0$. This algebraically
proves the geometrically intuitive conclusion made above: the
perturbed orbits are close to the $\theta' = {\rm const}$ lines.

Using the approximations (\ref{eq:energy_prime_approx})  and
(\ref{eq:orbit_shape}) for the energy and orbit shape, the integrals
(\ref{eq:dissipation_integral}) and (\ref{eq:period_integral}) can
be evaluated up to the first order in $\beta$. The details are given
in the Appendix~\ref{appendix_integrals perturbed_orbits}.
Substituting the results into Eq.~(\ref{eq:genericEvst}) we get a
differential equation
\begin{equation}\label{eq:approximate_diifeq}
\frac{d\varepsilon}{dt} = -\alpha \left(
 \frac{\partial\varepsilon_0}{\partial\theta}
  \right)^2_{\theta = \theta_0(\varepsilon)} + \  {\mathcal O}(\beta^2)
  \ .
\end{equation}
Up to the first order terms in $\beta$ this is the same equation as
one would have for magnetic switching in the unperturbed case with
$h_{\perp} = 0$. We conclude that in the rotated coordinates one can
approximate the switching time by the expression for the unperturbed
case. The latter \cite{kikuchi:1956, uesaka:2002} is reviewed in
Appendix~\ref{appendix_unperturbed_swithing_time} and gives the
switching time as a function of the starting angle $\theta_{in}$ and
the cut-off angle $\theta_{sw}$. To find the time $\tau_2$ spent by
the moment in Stage II we just have to set $\theta_{in}$ to be the
value of $\theta'$ at the end of Stage I, and $\theta_{sw}$ to be
the value of $\theta'$ at the selected switching moment given by
$\varepsilon = 0$. Thus $\theta_{in} =
\theta'(n_x(\tau_h),n_y(\tau_h))$ and $\theta_{sw} = \theta'(\theta
= \pi/2, \phi = \pi/2) = \pi/2$. The substitution is performed in
Appendix~\ref{appendix_unperturbed_swithing_time} and gives
\begin{eqnarray}
 \nonumber
 \tau_2  &=&  \frac{1}{2\alpha} \left\{
 \frac{1}{h_0 - \omega_0}
 \ln \left(
   \frac{h_0 - \omega_0 \cos\theta_{in}}{h_0(1-\cos\theta_{in})}
 \right) \right. -
 \\ \label{eq:unperturbed_switcing_time_sepcialized}
 && - \left.
 \frac{1}{h_0 + \omega_0}
 \ln \left(
   \frac{h_0 - \omega_0 \cos\theta_{in}}{h_0(1+\cos\theta_{in})}
 \right)
 \right\}
\end{eqnarray}

\subsection{Total switching time}
The total switching $\tau_m$ time is given by the sum of the
contributions from Stages I and II and equals $ \tau_m = \tau_h +
\tau_2$. To use the expression
(\ref{eq:unperturbed_switcing_time_sepcialized}) for $\tau_2$ we
need the value of $\theta_{in}$. This angle is given by the distance
between the endpoint of Stage I $\{n_x(\tau_h),n_y(\tau_h) \}$ and
the position of the energy maximum point $X$ given by
$(-h_{\perp}/(h_0-\omega_0),0)$ on a unit sphere. Since the point
$\{n_x(\tau_h),n_y(\tau_h)\}$ and the point $X$ are both close to
$+z$, we can approximately write
\begin{equation}\label{eq:theta_in}
\theta_{in} = \sqrt{
 (n_x(\tau_h) + h_{\perp}/(h_0-\omega_0))^2 + n_y(\tau_h)^2
 }
\end{equation}
A long expression for $\theta_{in}$ can be obtained by substituting
the formulae~(\ref{eq:nxny_at_tauh}) into the equation above.

As it was already discussed, the expressions (\ref{eq:nxny_at_tauh})
represent $n_x$ and $n_y$ as a sum of two terms, where the first
decreases and the second increases with time. Calculations are
substantially simplified when the increasing term dominates, which
turns out to be true in a large part of the parameter space. Even
for small values of $\alpha$ this is guaranteed for $1/(\omega_0 +
h_0) \ll \sqrt{\tau_h/h_0}$ or
\begin{equation}\label{eq:condition_second_term_domination}
\omega_0 \tau_h \gg \frac{\omega_0 h_0}{(\omega_0 + h_0)^2}
\end{equation}
Since we already assumed that $\omega_0 \tau_h \gg 1$, and the right
hand side of (\ref{eq:condition_second_term_domination}) is always
less than 1/4, this inequality is automatically satisfied whenever
we can use the steepest descent approximation
(\ref{eq:nxny_at_tauh}) for Stage I.

Next, we assume that one can also ignore $h_{\perp}/(h_0 -
\omega_0)$ in the first term of Eq.~(\ref{eq:theta_in}) compared
with $n_x(\tau_h)$ given by the dominant term of
(\ref{eq:nxny_at_tauh}). This condition is satisfied for $1/(h_0 -
\omega_0) \ll \sqrt{\tau_h/h_0}$ or
\begin{equation}\label{eq:condition_ignoring_distance_to_X}
\omega_0 \tau_h \gg \frac{\omega_0 h_0}{(h_0 - \omega_0)^2}
\end{equation}

As a result, leaving only the dominant terms we obtain
\begin{equation}\label{eq:theta_in_dominant}
\theta_{in} \approx h_{\perp} \sqrt{\frac{\pi\tau_h}{h_0}} \
e^{\alpha \Delta\varphi}
\end{equation}
Using the smallness of $\theta_{in}$, we approximate
$\sin\theta_{in} \approx \theta_{in}$ and $\cos\theta_{in} \approx
1$ in the expression
(\ref{eq:unperturbed_switcing_time_sepcialized}) and rewrite
$\tau_2$ as
\begin{eqnarray}
 \nonumber
 \tau_2  &=&  \frac{1}{2\alpha} \left\{
 \frac{1}{h_0 - \omega_0}
 \ln \left(
   \frac{2(h_0 - \omega_0)}{h_0 \theta_{in}^2}
 \right) \right. -
 \\ \nonumber
 && - \left.
 \frac{1}{h_0 + \omega_0}
 \ln \left(
   \frac{h_0 - \omega_0}{2 h_0}
 \right)
 \right\} =
 \\ \label{eq:tau2approx}
 &=& \frac{\ln(1/\theta_{in}^2)}{2\alpha(h_0 - \omega_0)} + \tau_R \ ,
\end{eqnarray}
where
\begin{eqnarray}
 \nonumber
 \tau_R(\alpha,h_0,\omega_0)  &=&  \frac{1}{2\alpha} \left\{
 \frac{1}{h_0 - \omega_0}
 \ln \left(
   \frac{2(h_0 - \omega_0)}{h_0}
 \right) \right. -
 \\ \label{eq:def_R}
 &&
 - \left.
 \frac{1}{h_0 + \omega_0}
 \ln \left(
   \frac{h_0 - \omega_0}{2 h_0}
 \right)
 \right\}
\end{eqnarray}
is a time interval independent of $\tau_h$ and $h_{\perp}$.

Substituting $\theta_{in}$ from Eq.~(\ref{eq:theta_in_dominant})
into Eq.~(\ref{eq:tau2approx}), we produce the first principal
result of our paper, a formula for the switching time
\begin{eqnarray}\label{eq:tau_m_derived}
\tau_m &=& \tau_h + \frac{\ln[h_0/\pi h_{\perp}^2 \tau_h] - \alpha
\Delta\phi(\tau_h)}{2 \alpha(h_0 -\omega_0)} + \tau_R
 \\
 \nonumber
 &=& \frac{3 h_0 + \omega_0}{4 h_0} \tau_h + \frac{\ln[h_0/\pi
h_{\perp}^2 \tau_h]}{2 \alpha(h_0 -\omega_0)} + \tau_R
(\alpha,h_0,\omega_0) \ .
\end{eqnarray}

The obtained $\tau_m(\tau_h)$ dependence indeed has a minimum. It is
reached at the optimal field sweep time
\begin{equation}\label{eq:optimal_tauh}
\tau_h^* = \frac{1}{2 \alpha (h_0 - \omega_0)}\frac{4 h_0 }{3 h_0 +
\omega_0}
\end{equation}
that is independent of the bias field. This formula is our second
main result. The independence of $\tau_h^*$ of $h_{\perp}$ is a
result of the logarithmic dependence in the second term of
(\ref{eq:tau_m_derived}) and ultimately stems from the logarithmic
dependence of $\tau_2$ on the initial deviation angle.

The minimal switching time $\tau_m(\tau_h^*)$ corresponding to the
optimal field sweep time equals to
$$
\tau_m(\tau_h^*) = \frac{1 +
 \ln\left(
   \frac{\alpha (h_0 - \omega_0)(3 h_0 + \omega_0)}{2 \pi h_{\perp}^2}
 \right)}{2\alpha (h_0 - \omega_0)}  + \tau_R \ .
$$
It does depend on $h_{\perp}$, which is a quite natural since the
initial deviation from the easy axis is controlled by the bias
field.

Expression (\ref{eq:tau_m_derived}) has its limits of applicability
discussed in the next section. In particular, it is not applicable
for small $\tau_h$ where the steepest descent approximation
(\ref{eq:stationary_phase_SandC}) is invalid. Nevertheless, one can
easily calculate $\tau_m(0)$ since in this case there is no motion
in Stage I, and, according to Figs.~\ref{fig:equilibria} and
\ref{fig:rotated_frame}, the initial angle for Stage II is simply
$$
\theta_{in} = \theta_u + \theta_* =  \frac{h_{\perp}}{h_0 +
\omega_0} + \frac{h_{\perp}}{h_0 - \omega_0} \ .
$$
Using this value of $\theta_{in}$ in (\ref{eq:tau2approx}) we get
\begin{equation}\label{eq:tauM_at_zero_tauH}
\tau_m(0) = \frac{1}{\alpha (h_0 - \omega_0)}
 \ln\left(
   \frac{h_0^2 - \omega_0^2}{2 h_0 h_{\perp}}
 \right) + \tau_R \ .
\end{equation}
The drop of switching time from $\tau_h = 0$ to the minimal value
is given by a formula
$$
\tau_m(0) - \tau_m(\tau_h^*) = \frac{\ln \left(
    \frac{\pi (h_0 - \omega_0)(h_0 + \omega_0)^2}{2 \alpha h_0^2 (3 h_0 + \omega_0)}
  \right) - 1}{2 \alpha (h_0 - \omega_0)} \ .
$$
Note that this difference is again independent of the bias field
$h_{\perp}$, as long as the approximation (\ref{eq:tau_m_derived})
is valid. The fractional change $(\tau_m(0) - \tau_m(\tau_h^*)
)/\tau_m(0)$ will depend on the bias field, being an increasing
function of $h_{\perp}$.

\section{Validity regions of the analytic
approximation}\label{sec:validity}

A number of approximations were made in our derivation and the final
expressions can only be used in the region on their validity.

The approximations made in our treatment of Stage I are as follows.

(a) The steepest descent method employed to evaluate the integrals
(\ref{eq:defSandC}) has to be sufficiently accurate. The required
conditions (see (\ref{eq:appendix_steepest_descent_conditions}) in
Appendix~\ref{appendixI}) read
\begin{equation} \label{eq:steepest_descent_validity}
\sqrt{\frac{\tau_h}{h_0}} \gg
  \frac{1}{h_0 - \omega_0} >  \frac{1}{h_0 + \omega_0} \ ,
\end{equation}
where the second inequality holds automatically.

(b) Next, we assumed that the inequality
$$
\frac{h_{\perp}}{h_0 + \omega_0} \ll
 h_{\perp} \sqrt{\frac{\pi\tau_h}{h_0}}
$$
is satisfied, allowing one to neglect the first terms on the right
hand sides of Eqs.~(\ref{eq:nxny_at_tauh}). However, this
requirement is not new because it is already contained in
(\ref{eq:steepest_descent_validity}).

(c) In order to make approximations in Eq.~(\ref{eq:theta_in}) we
required the inequality (\ref{eq:condition_second_term_domination})
to hold. This inequality is also contained in
(\ref{eq:steepest_descent_validity}) and does not add new
conditions.

(d) Finally, the inequality (\ref{eq:stageI_validity_condition})
should be satisfied to ensure small deviation of $\bf n$ from $+z$.

Overall, the inequalities
\begin{equation}\label{eq:summary_requirements_StageI}
\frac{h_{\perp}}{h_0 - \omega_0} \ll h_{\perp}
\sqrt{\frac{\tau_h}{h_0}} \ll  e^{-\alpha \Delta \varphi(\tau_h)} <
1
\end{equation}
summarize the requirements for Stage I. For our treatment of Stage
II we assumed the following.

(e) The Gilbert damping constant should be small, $\alpha \ll 1$, to
be able to use the orbit averaged equation of motion
(\ref{eq:genericEvst}).

(f) The parameter describing the orbit deformation
(\ref{eq:orbit_shape}) should be small, $\beta = h_{\perp}/(h_0 -
\omega_0) \ll 1$. But this inequality follows from
(\ref{eq:summary_requirements_StageI}) and thus brings no additional
restrictions.

The requirements (\ref{eq:summary_requirements_StageI}) discussed
above can be equivalently presented as conditions on $\tau_h$ that
have to be satisfied at fixed bias field $h_{\perp}$. In this form
they read
\begin{equation}\label{eq:tauHconditions}
\frac{h_0}{(h_0 - \omega_0)^2} \ll \tau_h \ll \tau_h^{(+)} \ ,
\end{equation}
where $\tau_h^{(+)}$ is a solution of
$$
\sqrt{\frac{\tau_h}{h_0}}
 e^{ -\alpha \Delta\varphi(\tau_h) } = \frac{1}{h_{\perp}} \ .
$$

We can now check when does the optimal field sweep time $\tau_h^*$
lie in the region of validity of our approximation. Using our result
(\ref{eq:optimal_tauh}) we can write
$$
\tau_h^*  \approx
 \frac{1}{2 \alpha (h_0 - \omega_0)} \ ,
$$
and substitute it into the requirement
(\ref{eq:steepest_descent_validity}). We get
\begin{equation*}
\alpha \ll \frac{h_0 - \omega_0}{2 h_0} \ .
\end{equation*}
The right hand side of this inequality is always smaller than unity,
thus it automatically implies $\alpha \ll 1$.

We also have to satisfy the condition
(\ref{eq:stageI_validity_condition}) at $\tau_h = \tau_h^*$. For the
exponent $\alpha\Delta\phi$ one can write
$$
\alpha \Delta\varphi (\tau_h^*) =
 \alpha\frac{(h_0 - \omega_0)^2}{4 h_0} \tau_h^*
 \approx \frac{h_0 -\omega_0}{8h_0} \leq \frac{1}{8} \ll 1 \ ,
$$
and thus condition (\ref{eq:stageI_validity_condition}) simplifies
to
$$
h_{\perp}\sqrt{\frac{\tau_h^*}{h_0}}
 \ll 1
$$
The above inequalities on $\alpha$ and $h_{\perp}$ can be combined
into a single requirement
\begin{equation}\label{eq:alpha_requirement}
\frac{h_{\perp}^2}{h_0(h_0-\omega_0)} \ll \alpha \ll \frac{h_0 -
\omega_0}{2 h_0} \ .
\end{equation}
If the inequalities (\ref{eq:alpha_requirement}) are satisfied, the
optimal sweep time $\tau_h^*$ occurs inside the interval
(\ref{eq:tauHconditions}) and can be calculated using the formula
(\ref{eq:optimal_tauh}).

\begin{figure}[t]
\center
\includegraphics[width=0.37\textwidth]{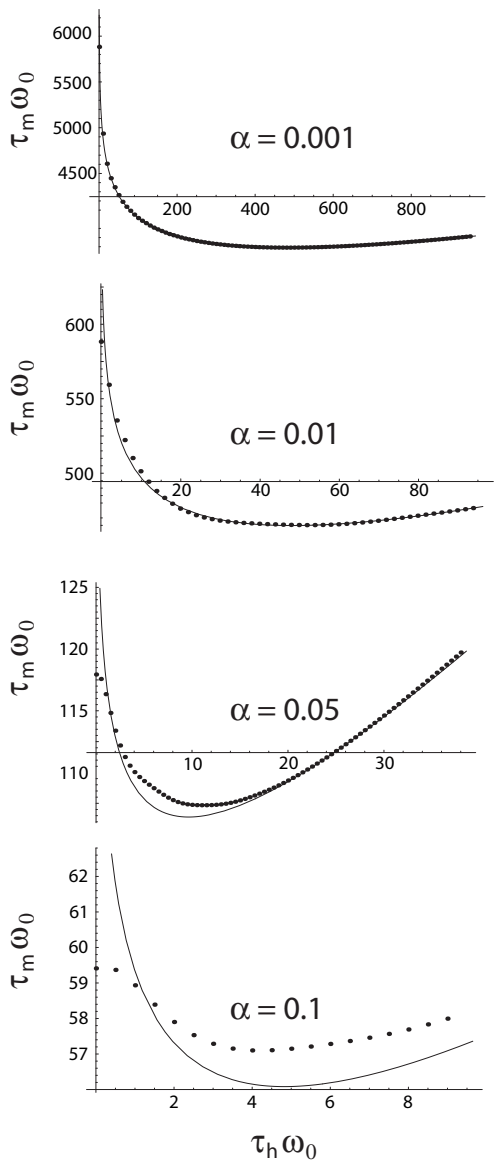}
\caption{Dependencies $\tau_m(\tau_h)$ calculated for $h_0 = 2.2 \
\omega_0$, $h_{\perp} = 0.001 \ \omega_0$, and variable $\alpha$
indicated on each panel. As $\alpha$ increases, the theoretical fit
gets poorer due to the violation of the strong inequality
(\ref{eq:alpha_requirement}).}
 \label{fig:alpha_family}
\end{figure}

\section{Comparison of analytic and numeric results}

Comparisons of the analytic approximation and exact numeric results
were performed in Ref.~\onlinecite{bazaliy:2010} and shown a very
good agreement between the two.

First, we compared the numerically calculated switching times with
the expression (\ref{eq:tau_m_derived}). When the inequalities
(\ref{eq:alpha_requirement}) were well satisfied, the quality of
approximation was very good (Fig.~\ref{fig:alpha_family}). As one
approached the limits of the approximation's validity by, e.g.,
increasing $\alpha$, the errors grew larger.

\begin{figure}[t]
\center
\includegraphics[width=0.37\textwidth]{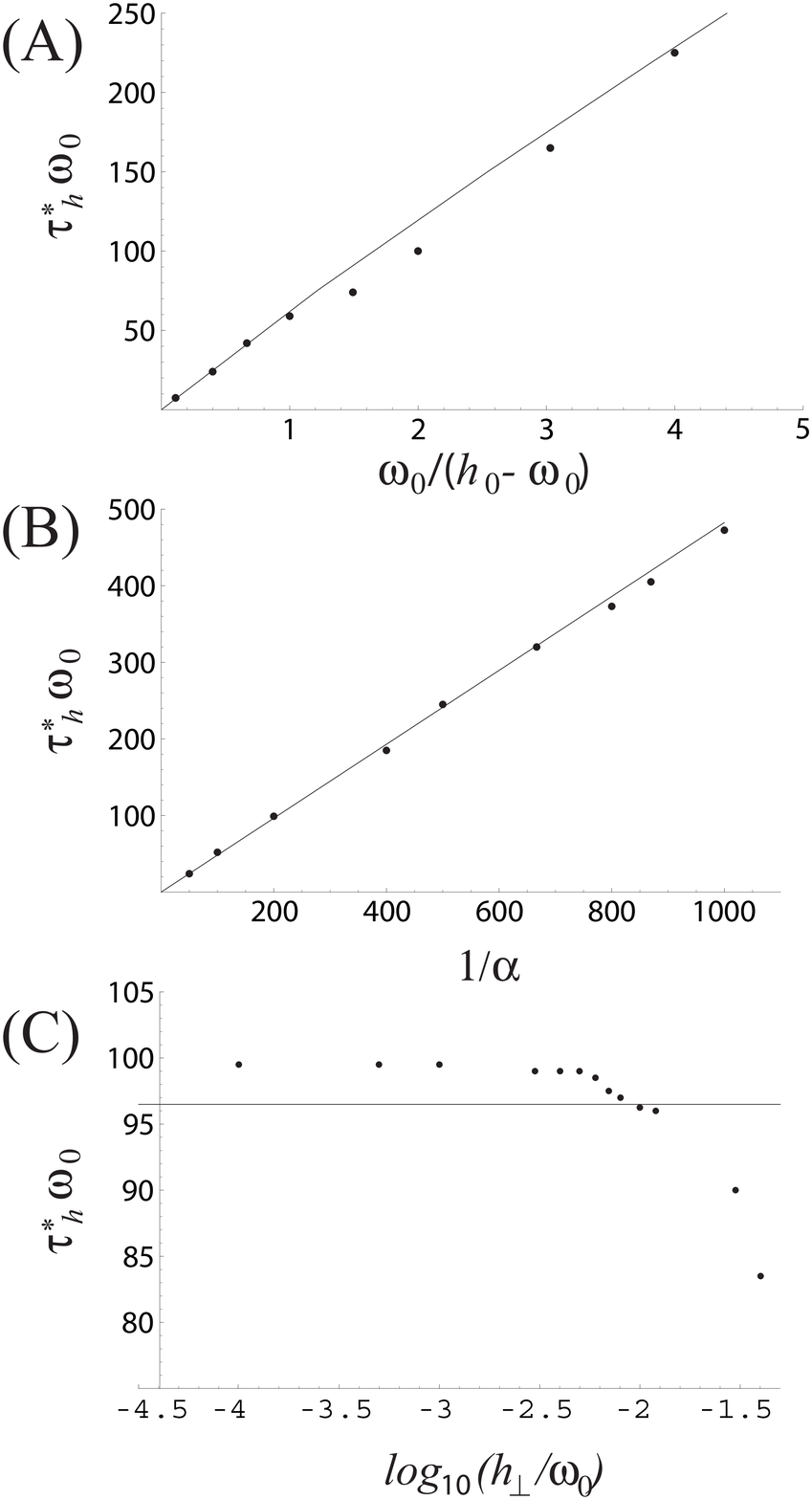}
\caption{Numeric (points) and approximate analytical (solid lines)
dependencies of the optimal field sweep time $\tau_h^*$ on the
system parameters. (A) fixed $\alpha$ and $h_{\perp}$, (B) fixed
$h_0/\omega_0$ and $h_{\perp}$, (C) fixed $h_0/\omega_0$ and
$\alpha$. When not varied, the parameter values are $h_0 = 2.2
\omega_0$, $\alpha = 0.01$, $h_{\perp} = 0.005$. }
 \label{fig:taustar}
\end{figure}

Second, the numeric results for optimal field sweep time
$\tau_h^{*}$ were compared with the formula (\ref{eq:optimal_tauh}).
The correspondence was generally good (Fig.~\ref{fig:taustar}),
although some visible deviations existed. They were attributed to
the fact that the accuracy of the determination of $\tau_h^*$ is
lowered by a flat shape of the $\tau_m(\tau_h)$ curve minimum.
Because of the shallow minimum, small errors in $\tau_m$ produce
much larger errors in $\tau_h^*$.

In general, the analytic expression approximated the
$\tau_m(\tau_h)$ dependence up to a 10\% accuracy in a surprisingly
wide range of parameters. Such accuracy is certainly sufficient for
the estimates related to the device design.

When $\tau_h$ is outside of the validity region of the results
(\ref{eq:tau_m_derived}) and (\ref{eq:optimal_tauh}), exact
expressions (\ref{eq:defSandC}) and (\ref{eq:theta_in}) for the
integrals and the initial angle $\theta_{in}$ can be used. As long
as the deviation from the $+z$ direction in Stage I remains small,
they provide a good approximation for $\tau_m$. Consider for example
the case of small $\tau_h$. Approximation (\ref{eq:tau_m_derived})
does not work for $\tau_h \to 0$ predicting an infinite increase of
$\tau_m$, while the actual limit $\tau_m(0)$ is finite and given by
formula (\ref{eq:tauM_at_zero_tauH}). It was checked that using the
exact expressions (\ref{eq:appendix_exact_J}) for the integrals $C$
and $S$ and the formula (\ref{eq:theta_in}) for $\theta_{in}$, one
can obtain an excellent agreement between the two-stage theory and
the no-approximation numeric simulations in this limit.

\section{Physical picture of the ballistic-assisted switching}

We now discuss the physical reason for the minimum of the function
$\tau_m(\tau_h)$ which was identified in
Ref.~\onlinecite{bazaliy:2010} as the contribution of ballistic
switching. We start by noticing that the bias field has two roles in
the switching process. First, it provides the initial deviation from
the easy axis. Second, it alters the equations of motion for ${\bf
n}(t)$. The first role of $h_{\perp}$ manifests itself in our
formulae in two ways: by providing the first terms in expressions
(\ref{eq:nxny_at_tauh}) and by introducing the term
$h_{\perp}/(h_0-\omega_0)$ into the formula (\ref{eq:theta_in}). In
our derivation of the switching time expression
(\ref{eq:tau_m_derived}) we have found that both contributions are
negligible. Therefore within our approximation only the second role
of the bias field is important.

\begin{figure}[t]
 \center
\includegraphics[width = 0.25 \textwidth]{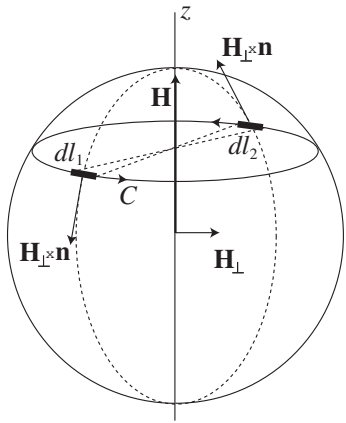}
\caption{Average ballistic contribution of the bias field. Vector
${\bf n}(t)$ orbits around a parallel circle $C$ on a unit sphere.
The torque due to $h_{\perp}$ pushes $\bf n$ along the meridians of
the sphere. In constant switching field the torque contributions
from the diametrically opposed elements $dl_1$ and $dl_2$ cancel
each other. For variable ${\bf H}(t)$ the cancelation does not
happen (see text). }
 \label{fig:ballistic_contribution_averaging}
\end{figure}

Let us proceed by discussing this role qualitatively. Recall that in
the absence of anisotropy and other fields the torque ${\bf
H}_{\perp} \times M_0{\bf n}$ due to the bias field would rotate the
unit vector $\bf n$ from $+ z$ to $- z$ along a meridian of the unit
sphere (dashed line in
Fig.~\ref{fig:ballistic_contribution_averaging}), in a ``ballistic''
or ``precessional'' fashion. In our case a weak bias field is
applied on top of the strong uniaxial anisotropy and switching
field, which induce a fast orbital motion of vector ${\bf n}(t)$
along the parallel circles (line $C$ in
Fig.~\ref{fig:ballistic_contribution_averaging}). The bias field
still attempts to move $\bf n$ along the meridians, but now its
action has to be averaged over the period of orbital motion. As
illustrated in Fig.~\ref{fig:ballistic_contribution_averaging}, in
constant fields ${\bf H} || \hat z$ averaging gives zero due to the
cancelation of the contributions from  the diametrically opposed
infinitesimal intervals $dl_1$ and $dl_2$ of equal lengths. This way
ballistic contribution of the bias field is quenched. However, the
contribution of ${\bf H}_{\perp}$ does not average to zero for a
variable switching field ${\bf H}(t)$. In this case the velocity of
$\bf n$ changes along the orbit, the times spent in the intervals
$dl_1$ and $dl_2$ are different, and the contributions of the two do
not cancel each other. We conclude that in the presence of a time
dependent field ${\bf H}(t)$ ballistic contribution of the
perpendicular bias field is partially recovered.

\begin{figure}[t]
\center
\includegraphics[width = 0.4\textwidth]{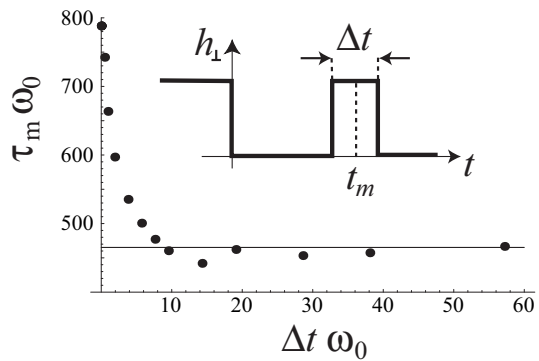}
\caption{Numerically calculated witching time with pulsed bias
field. The parameters are set to $h_0 = 2.2 \omega_0$, $h_{\perp} =
0.001 \omega_0$, $\alpha = 0.01$ (cf. Fig.~\ref{fig:alpha_family}).
Field sweep time is fixed at $\tau_h = 50/\omega_0$, close to the
optimal sweep time $\tau_h^* = 48.2/\omega_0$. The time dependence
of the pulsed bias field $h_{\perp}(t)$ is shown in the inset. As
the width of the pulse $\Delta t$ approaches the theoretical target
of $2\sqrt{\tau_h/h_0} \approx 10/\omega_0$, the switching time
approaches the value obtained at $h_{\perp} = $ const (shown by a
horizontal line).  }
 \label{fig:pulse_Hp}
\end{figure}

Based on the discussion above, we may expect to find that the
largest contribution of the ballistic switching will happen when the
change of velocity is biggest. The measure of the velocity change is
$\Delta\omega/\omega$, where $\Delta\omega$ is the change of the
velocity as one precesses from the interval $dl_1$ to the interval
$dl_2$. We may estimate
$$
\frac{\Delta\omega}{\omega} \sim \frac{\dot \omega T}{\omega} \sim
\frac{\dot \omega}{\omega^2}
  \sim \frac{\ddot\phi}{\dot\phi^2} \ ,
$$
where $T = 2\pi/\omega$ is the instantaneous period. We see that the
velocity change estimate diverges at the point $\dot\phi = 0$. But
this is exactly the stationary phase point $t_m$ that gives the
largest contribution to the integrals (\ref{eq:defSandC}) in our
approximation. We have thus established a one-to-one correspondence
between the qualitative description of the ballistic contribution
and our derivation of the analytic approximation
(\ref{eq:tau_m_derived}). To illustrate the fact that the action of
$h_{\perp}$ is only important near the stationary phase point $t_m$,
we have performed a numeric experiment with pulsed bias field
$h_{\perp}(t)$ changing in time as shown in the inset of
Fig.~\ref{fig:pulse_Hp}. It is kept constant for $t < 0$, switched
off at $t = 0$, and then switched on again for a short interval of
time $\Delta t$, centered around the $t_m$ point. With the pulsed
bias field the ballistic contribution is only present during the
interval $\Delta t$. It follows from the steepest descent
calculation of Appendix~\ref{appendixI} that formulae
(\ref{eq:stationary_phase_SandC}) would be valid already for $\Delta
t \gtrsim 2\sqrt{\tau_h/h_0}$, and when this inequality is satisfied
our theory would predict the same switching time
(\ref{eq:tau_m_derived}) for pulsed and constant bias fields. The
results of the numeric experiment (Fig.~\ref{fig:pulse_Hp}) are
completely consistent with this prediction. As the pulse width
approaches the value of $2\sqrt{\tau_h/h_0}$, the switching time
drops to the value obtained earlier at constant $h_{\perp}$.

The minimum of the $\tau_m(\tau_h)$ function can be understood as
follows. Ballistic contribution helps to move vector $\bf n$ from $+
z$ to $- z$ and is thus responsible for the initial decrease of
$\tau_m$. As the sweep time grows larger, the change of the orbital
velocity during the precession period decreases and the ballistic
contribution averages out progressively better. The helping effect
of ballistic switching is lost and $\tau_m$ starts to increase as it
normally would.

Finally, we want to remark that ballistic contribution to switching
can be also viewed as a phenomenon complimentary to the magnetic
resonance and RF-assisted switching.\cite{thirion:2003, coffey:1998,
garcia:1998, bertotti:2001, rivkin:2006, zzsun:2006RF, scholz:2008}
In the case of RF-field application the external field ${\bf H}$ is
constant, while the bias field ${\bf H}_{\perp}(t)$ is
time-dependent. Here the contributions of the bias field torque on
the intervals $dl_1$ and $dl_2$ in
Fig.~\ref{fig:ballistic_contribution_averaging} do not cancel each
other because the torque itself changes with time. This, again,
leads to a nonzero average of the bias field contribution on a
precession orbit and creates a helping effect for the magnetic
switching process. From this point of view the magnetic resonance
and the time dependence of the axial switching field are two
different ways to achieve the same goal: a non-vanishing average
contribution of the bias field torque on an orbit.

\section{Conclusions}

We have identified and investigated the phenomenon of ballistic
contribution to the conventional magnetic switching by a
time-dependent field. An analytic approximation is derived for the
ballistic-assisted switching time in a constant perpendicular bias
field. It is also shown that, if practical, a constant bias can be
substituted by short pulse of bias field applied near the stationary
phase time point. Our results provide a convenient approximation for
the optimal field sweep time, an important parameter in the device
design. The expressions obtained in this study can be used as a
starting point for the investigations of the switching time in
granular media, where each grain can be modeled by a single moment
and bias field is produced by the other grains or by the spread of
grain orientations.

\section{Acknowledgments}
Ya.~B. Bazaliy is grateful to B. V. Bazaliy for illuminating
discussions. This work was supported by the NSF grant DMR-0847159.

\appendix

\section{Steepest descent approximation for $C(t)$ and $S(t)$}\label{appendixI}
Here we evaluate the integrals (\ref{eq:defSandC})
\begin{eqnarray*}
S(\tau_h) &=& \int_0^{\tau_h} e^{\alpha\varphi(t)} \sin\varphi(t) dt
 \\
C(\tau_h) &=& \int_0^{\tau_h} e^{\alpha\varphi(t)} \cos\varphi(t) dt
\end{eqnarray*}
with the phase $\varphi(t)$ given by a real quadratic function
$$
\varphi = (\omega_0 + h_0) t - h_0 \frac{t^2}{\tau_h} \ .
$$
The integrals in question can be obtained from the real and
imaginary parts of a complex integral
\begin{eqnarray}\label{eq:appendix_definitin_I}
I &=& \int_0^{\tau_h} e^{-\mu\varphi(t)}  dt = C - iS \ ,
 \\ \nonumber
\mu &=& i - \alpha \ .
\end{eqnarray}
By completing the square we can rewrite
$$
\varphi(t) = \frac{(h_0 + \omega_0)^2}{4 h_0}\tau_h -
\frac{h_0}{\tau_h} (t - t_m)^2 \ ,
$$
where $t_m = \tau_h(\omega_0 + h_0)/2h_0$ is the point of maximum
phase, $0 < t_m < \tau_h$. Now
\begin{eqnarray}
 \label{eq:appendix_definition_J}
I &=& e^{\mu \frac{(h_0 + \omega_0)^2}{4 h_0} \tau_h} \cdot J \ ,
 \\ \nonumber
J &=&  \int_0^{\tau_h}
 e^{-\mu (h_0/\tau_h)(t-t_m)^2} dt \ .
\end{eqnarray}
Changing variables to $z = \sqrt{\mu h_0/\tau_h}(t-t_m)$, we can
write down $J$ as
$$
J = \sqrt{\frac{\tau_h}{\mu h_0}} \int_{\Gamma} e^{- z^2} dz \ ,
$$
where $\Gamma$ is a straight line in the complex plane going between
the points $z_1 = -\sqrt{\mu h_0/\tau_h} \ t_m$ and $z_2 = \sqrt{\mu
h_0/\tau_h} \ (\tau_h - t_m)$. Due to our definition of $z$, line
$\Gamma$ crosses the complex zero point.

Since the integrand of $J$ is a regular function, integration can be
performed along any contour connecting $z_1$ and $z_2$. The integral
can be expressed in terms of the error function of complex variable
${\rm Erf}(z) = (2/\sqrt{\pi}) \int_0^z \exp(-z^2) dz$ as
\cite{abramovitz_stegun}
\begin{equation}\label{eq:appendix_exact_J}
J = \sqrt{\frac{\tau_h}{\mu h_0}} \frac{\sqrt\pi}{2}
 \left( {\rm Erf}(z_1) + {\rm Erf}(z_2) \right)
\end{equation}
The above in an exact formula. The steepest descent approximation
corresponds to the case of large absolute values $|z_1| \gg 1$,
$|z_2| \gg 1$. Due to the smallness of $\alpha$ one has $|\mu|
\approx 1$, so these conditions translate to
$$
\sqrt{\frac{h_0}{\tau_h}} t_m \gg 1
 \ , \quad
 \sqrt{\frac{h_0}{\tau_h}} (\tau_h - t_m) \gg 1  \ ,
$$
or equivalently
\begin{equation}\label{eq:appendix_steepest_descent_conditions}
\frac{(h_0 + \omega_0)^2}{4 h_0} \tau_h \gg 1
 \ , \quad
\frac{(h_0 - \omega_0)^2}{4 h_0} \tau_h \gg 1
\end{equation}
Using \cite{abramovitz_stegun} ${\rm Erf}(z) \to 1$ for $|z| \to
\infty$ we find the approximation
$$
J \approx \sqrt{\frac{\pi\tau_h}{\mu h_0}} \approx
 \sqrt{\frac{\pi\tau_h}{h_0}} e^{i\pi/4}
 \left(
 1 - \frac{i\alpha}{2}
 \right) \ ,
$$
where we have also expanded in small $\alpha$. Substituting this
back into (\ref{eq:appendix_definition_J}) we get
\begin{equation}\label{eq:appendix_approx_I}
I \approx e^{\alpha \varphi_m} \sqrt{\frac{\pi\tau_h}{h_0}}
 e^{i(\pi/4 -\varphi_m)} \left( 1 - \frac{i\alpha}{2} \right)
\end{equation}
where $\varphi_m = \varphi(t_m)$. The real and imaginary parts of
$I$ give $C$ and $S$ according to
Eq.~(\ref{eq:appendix_definitin_I})
\begin{eqnarray*}
C &\approx& e^{\alpha\phi_m} \sqrt{\frac{\pi\tau_h}{h_0} } \left[
 \cos\left(\phi_m - \frac{\pi}{4}\right)
 - \frac{\alpha}{2} \sin\left(\phi_m - \frac{\pi}{4}\right)
 \right]
  \ ,
 \\
S &\approx& e^{\alpha\phi_m} \sqrt{\frac{\pi\tau_h}{h_0} } \left[
 \sin\left(\phi_m - \frac{\pi}{4}\right)
 + \frac{\alpha}{2} \cos\left(\phi_m - \frac{\pi}{4}\right)
 \right] \ .
\end{eqnarray*}

\section{Energy in rotated coordinates}\label{appendix_energy_rotated_coordinates}
The relationship between the projections of $\bf n$ and $\bf h$ in
the primed and original coordinate system are given as
\begin{eqnarray*}
n_z &=& n'_z \cos\theta_* + n'_x \sin\theta_*
 \\
 &=& \cos\theta'\cos\theta_* + \sin\theta'\cos\phi'\sin\theta_*
 \\
n_x &=& -n'_z\sin\theta_* + n'_x\cos\theta_* =
 \\
 &=& -\cos\theta'\sin\theta_* + \sin\theta'\cos\phi'\cos\theta_*
 \\
n_y &=& n'_y
\end{eqnarray*}
and
\begin{eqnarray*}
h'_z &=& h \cos\theta_* - h_{\perp}\sin\theta_*
 \\
h'_x &=& h \sin\theta_* + h_{\perp}\cos\theta_*
\end{eqnarray*}
One can now rewrite Eq.~(\ref{eq:uniaxial_total_energy_1}) through
the angles $(\theta',\phi')$
\begin{eqnarray}
 \nonumber
\varepsilon &=& -\frac{\omega_0}{2} {n'_z}^2 - n'_x h'_x - n'_z h'_z
=
 \\ \nonumber
 &=& -\frac{\omega_0}{2}(\cos\theta'\cos\theta_* +
 \sin\theta'\cos\phi'\sin\theta_*)^2
 \\ \label{eq:energy_prime_exact}
 &&- h(\cos\theta'\cos\theta_* + \sin\theta'\cos\phi'\sin\theta_*)
 \\ \nonumber
 && - h_{\perp} (\sin\theta'\cos\phi'\cos\theta_*
  - \cos\theta'\sin\theta_*)
\end{eqnarray}
The above is the exact expression. We are looking at the case $h =
-h_0$ and $h_{\perp} \to 0$. Up to the first order in $h_{\perp}$
$$
\sin\theta_* \approx  \frac{h_{\perp}}{h_0 - K} ,
 \quad
\cos\theta_* \approx  1 \ .
$$
Expanding the energy up to the first order in $h_{\perp}$ we get
\begin{equation}\label{eq:energy_prime_approx_appendix}
\varepsilon \approx  \varepsilon_0(\theta')
 + \beta \varepsilon_1(\theta',\phi')
\end{equation}
where the small parameter is
\begin{equation}\label{eq:defbeta_appendix}
\beta = \frac{h_{\perp}}{h_0 - \omega_0}
\end{equation}
and
\begin{eqnarray}
 \nonumber
\varepsilon_0(\theta') &=& -\frac{\omega_0}{2} \cos^2\theta'
 + h_0 \cos\theta'
 \\ \label{eq:appendix_varepsilon1}
\varepsilon_1(\theta') &=& \omega_0 (1 -
\cos\theta')\sin\theta'\cos\phi'
\end{eqnarray}
The first term in the expansion (\ref{eq:energy_prime_approx}) is
the energy unperturbed by the bias field, evaluated at the new polar
angle $\theta'$.

\section{Integrals along the perturbed orbits}\label{appendix_integrals perturbed_orbits}

First, we calculate the approximate value of $|\partial
\varepsilon/\partial {\bf n}|$. Taking the identity
$$
\left| \frac{\partial \varepsilon}{\partial {\bf n}}\right| =
  \sqrt{
  \left( \frac{\partial \varepsilon}{\partial \theta'} \right)^2 +
  \frac{1}{\sin^2\theta'}
  \left( \frac{\partial \varepsilon}{\partial \phi'} \right)^2
  }
$$
and expanding in small $\beta$ up to the first order we find
\begin{equation}\label{eq:energy_gradient_prime_approx}
\left| \frac{\partial \varepsilon}{\partial {\bf n}}\right|
 = \frac{\partial\varepsilon_0}{\partial\theta'} +
 \beta \frac{\partial\varepsilon_1}{\partial\theta'} + \ldots
\end{equation}

Next, we need the element of orbit length $|dn|$. Using $|dn| =
\sqrt{\sin^2\theta d\phi'^2 + d\theta'^2}$ and calculating up to the
first order in $\beta$ we get
\begin{equation} \label{eq:dn_approx_appendix}
|dn| = \left(\sin\theta_0 - \beta \frac{\cos\theta_0 \
\varepsilon_1(\theta_0,\phi')} {(d\varepsilon_0/d\theta)|_{\theta =
\theta_0(\varepsilon)}} \right) d\phi'
\end{equation}
Using the form of $\varepsilon_1$ (\ref{eq:appendix_varepsilon1}) we
will rewrite it as
$$
|dn| = \left(\sin\theta_0 - \beta A(\varepsilon)\cos\phi' \right)
d\phi'
$$
We will further use a notation
$$
\varepsilon_1(\theta_0(\varepsilon),\phi') = B(\varepsilon)\cos\phi'
$$

To perform the integrals (\ref{eq:dissipation_integral}) and
(\ref{eq:period_integral}) we use the expansions
(\ref{eq:energy_gradient_prime_approx}) and
(\ref{eq:dn_approx_appendix}), and, expanding up to the first order
in $\beta$, get
\begin{eqnarray*}
\oint_{\Gamma}  \left| \frac{\partial E}{\partial {\bf n}} \right|
dn &=& \int_0^{2\pi} \left[
 \frac{\partial \varepsilon_0}{\partial\theta} \sin\theta_0 + \right.
 \\
 &+& \left. \beta \left(
   \frac{\partial B}{\partial\theta}\sin\theta_0 -
   \frac{\partial \varepsilon_0}{\partial\theta} A \right)
   \cos\phi' + \ldots
 \right]  d\phi'
\end{eqnarray*}
The integral of first order term in $\beta$ vanishes and we get
$$
\oint_{\Gamma} \left| \frac{\partial E}{\partial {\bf n}} \right| dn
 = 2\pi \frac{\partial \varepsilon_0}{\partial\theta}
  \sin\theta_0 + {\mathcal O}(\beta^2)
$$
Perform a similar calculation for the integral
(\ref{eq:period_integral}) we get
$$
\oint_{\Gamma} \frac{dn}{\left|\partial \varepsilon/ \partial {\bf
n} \right|} = 2\pi \frac{\sin\theta_0}{\partial
\varepsilon_0/\partial\theta} + {\mathcal O}(\beta^2)
$$
According to (\ref{eq:genericEvst}) the differential equation on
$\varepsilon(t)$ reads
\begin{equation}\label{eq:appendix_approximate_diifeq}
\frac{d\varepsilon}{dt} = -\alpha \left. \left(
 \frac{\partial\varepsilon_0}{\partial\theta}
  \right)^2\right|_{\theta = \theta_0(\varepsilon)} + \ {\mathcal O}(\beta^2)
\end{equation}

\section{Switching time in the unperturbed case} \label{appendix_unperturbed_swithing_time}
The problem of switching time of a uniaxial particle in the absence
of perpendicular field was probably first solved by Kikuchi
\cite{kikuchi:1956} for the case of $H_z = 0$. The derivation was
generalized to arbitrary $H_z$ by many authors, in particular in the
appendix of Ref.~\onlinecite{uesaka:2002}. Here we re-derive this
result for the completeness of the presentation. In the case of
$h_{\perp} = 0$ one can find the switching time either by solving
Eq.~(\ref{eq:approximate_diifeq}) (same as
Eq.~(\ref{eq:appendix_approximate_diifeq})) truncated to zeroth
order, or by a direct inspection of the system (\ref{LLGtheta}),
(\ref{LLGphi}). Since $\varepsilon_0$ depends only on $\theta$, it
is enough to consider the first equation which reads
$$
\dot\theta = -\alpha\frac{\partial\varepsilon}{\partial\theta} =
 -\alpha(\omega_0\cos\theta + h )\sin\theta
$$
Integrating  we get
$$
-\alpha t = \int_{\theta_{in}}^{\theta_{sw}}
 \frac{d\theta}{\sin\theta(\omega_0\cos\theta + h)} \ .
$$
A variable change $x = \cos\theta$ gives
\begin{eqnarray}
 \nonumber
 t &=& \frac{1}{\alpha}\int_{x_{in}}^{x_{sw}}
 \frac{dx}{(1-x^2)(\omega_0 x + h)} =
 \\ \nonumber
 &=&  -\frac{1}{2\alpha} \left\{
 \frac{1}{h + \omega_0}
 \ln \left(
   \frac{1-x}{\omega_0 x + h}
 \right) \right. -
 \\ \label{eq:unperturbed_switcing_time_generic}
 && - \left.\left.
 \frac{1}{h - \omega_0}
 \ln \left(
   \frac{1+x}{\omega_0 x + h}
 \right)
 \right\}\right|_{x_{in}}^{x_{sw}}
\end{eqnarray}
In application to our problem $\theta_{sw} = \pi/2$ ($x_{sw} = 0$)
and $h = -h_0$ which gives
\begin{eqnarray}
 \nonumber
 t  &=&  \frac{1}{2\alpha} \left\{
 \frac{1}{h_0 - \omega_0}
 \ln \left(
   \frac{h_0 - \omega_0 \cos\theta_{in}}{h_0(1-\cos\theta_{in})}
 \right) \right. -
 \\ \label{eq:unperturbed_switcing_time_sepcialized_appendix}
 && - \left.
 \frac{1}{h_0 + \omega_0}
 \ln \left(
   \frac{h_0 - \omega_0 \cos\theta_{in}}{h_0(1+\cos\theta_{in})}
 \right)
 \right\}
\end{eqnarray}

\end{document}